\documentclass[preprint,showpacs,preprintnumbers,amsmath,amssymb]{revtex4}

\usepackage{amsmath}
\usepackage{epsf}
\usepackage{epsfig}
\begin{document}

\title{ The structure of the free energy surface of coarse-grained off-lattice protein models}
\author{ Ethem Akt\"{u}rk and  Handan Arkin }
\email{ handan@hacettepe.edu.tr  (corresponding-author)}
\affiliation{ Hacettepe University, Department of Physics Engineering
               Beytepe, Ankara, Turkey  }
\author{ Tar{\i}k \c{C}elik }
\affiliation{Turkish Academy of Sciences, Piyade Sok. No: 27, \c{C}ankaya,
Ankara, Turkey}
\begin{abstract}

We have performed multicanonical simulations of hydrophobic-hydrophilic heteropolymers with a simple effective, coarse-grained off-lattice model to study the
structure  and the topology of the  energy surface.
The  multicanonical method samples the whole rugged energy landscape, in particular
 the low-energy part, and enables one to better understand
the critical behaviors and visualize the
folding pathways of the considered protein model.


\end{abstract}

\maketitle

\section{Introduction}
 Predicting the proteins structure and the folding mechanism is an important goal in structural biology.
 Although the physical
principles are known, the complexity of proteins as being macromolecules
consisting of numerous atoms
makes an accurate analysis of the folding process of realistic
proteins extremely difficult. The configuration space of proteins
 presents a complex energy profile consisting of tremendous
number of local minima, barriers and further topological features.
Because of energy barriers, the commonly used thermodynamic
simulation techniques are not very efficient in sampling of a
rugged landscape. The simulated molecule often remains in its
starting wide microstate or move to a neighboring
 wide microstate, but in practise it will hardly reach the most stable one.
The system may be trapped in a basin for a long time,
which results in nonergodic behavior.
 On the other hand, the topography of the energy landscape,
 especially near the global minimum
 is of particular importance; this is because the potential energy surface
defines the behavior of the system. Consequently, a visualization
of the whole rugged landscape, covering the entire energy and
temperature ranges, would be helpful in developing methods to
allow one to survey the distribution of structures in
conformational space
 and also gives the details of the folding pathway. Such a goal can be
 achieved with the multicanonical approach.

\bigskip

 The problem of protein folding entails the study of a nontrivial dynamics along pathways embedded in a rugged energy landscape. Therefore, one of the  important aspects
in this field is studying simple
effective, coarse-grained models  that
allows  a more global view on the relationship between the
sequence of amino acid residues and the existence of a  funnel-like structure
along a pathway towards the
energy minimum in a rugged free-energy landscape~\cite{onuchic1}.

\bigskip

One of the most known  examples is the HP model of lattice
proteins,   which has been exhaustively
investigated~\cite{dill1,Larson}. In this model, only two types of
monomers are considered, with hydrophobic ($H$) and polar ($P$)
character. Chains on the lattice are self-avoiding to account for
the excluded volume. The only explicit interaction is between
non-adjacent but next-neighbored hydrophobic monomers. This
interaction of hydrophobic contacts is attractive to force the
formation of a compact hydrophobic core which is screened from the
(hypothetic) aqueous environment by the polar residues.
Statistical mechanics aspects of this model are being subject of
recent studies~\cite{iba1,hsu1,bj1}.

\bigskip

Another
off-lattice generalisation of the HP model is the AB model~\cite{ab1}, where the hydrophobic monomers are
labelled by $A$ and the polar or hydrophilic ones by $B$. The contact interaction is replaced by
a more realistic distance-dependent Lennard-Jones type of potential accounting for
short-range excluded volume repulsion and long-range interaction; the latter being attractive for $AA$ and
$BB$ pairs and repulsive for $AB$ pairs of monomers. An additional interaction accounts for the bending
energy of any pair of successive bonds. This model was first applied in two dimensions~\cite{ab1} and
generalized to three-dimensional AB proteins~\cite{irb1,irb2},
 with modifications by taking implicitly into account the additional
torsional energy contributions of each bond.
\bigskip

\section{The Model}

In this work,  we study  the effective off-lattice AB model of
 heteropolymers with $N$ monomers.
AB model as proposed in Ref.~\cite{irb1} has the energy function

\begin{eqnarray}
\label{irbaeck}
E&=&-\kappa_1\sum\limits_{k=1}^{N-2}{\bf b}_k\cdot{\bf b}_{k+1}-
\kappa_2\sum\limits_{k=1}^{N-3}{\bf b}_k\cdot{\bf b}_{k+2}+\nonumber\\
&&\hspace{3mm}4\sum\limits_{i=1}^{N-2}\sum\limits_{j=i+2}^NC(\sigma_i,\sigma_j)\left(\frac{1}{r_{ij}^{12}}
-\frac{1}{r_{ij}^6} \right),
\end{eqnarray}
where ${\bf b}_k$ is the bond vector between the monomers $k$ and $k+1$ with length unity. In Ref.~\cite{irb1}
different values for the parameter set ($\kappa_1$, $\kappa_2$) were tested and finally set to ($-1$, $0.5$)
as this choice provide both the  distributions for the angles between bond vectors ${\bf b}_k$ and ${\bf b}_{k+1}$
and  the torsion angles between the surface vectors ${\bf b}_k\times{\bf b}_{k+1}$
and ${\bf b}_{k+1}\times{\bf b}_{k+2}$ giving the  best agreement with the
distributions obtained for selected
functional proteins. Since ${\bf b}_k\cdot{\bf b}_{k+1}=\cos \vartheta_k$, the choice $\kappa_1=-1$ makes
the coupling between successive bonds ``antiferromagnetic''.
 The second term in Eq.~(\ref{irbaeck}) takes torsional
interactions into account.
 The third term contains  a pure Lennard-Jones potential, where the $1/r_{ij}^6$ long-range interaction
is attractive whatever types of monomers interact. The monomer-specific prefactor $C(\sigma_i,\sigma_j)$
only controls the depth of the Lennard-Jones valley:
\begin{equation}
\label{irbC}
C(\sigma_i,\sigma_j)=\left\{\begin{array}{cl}
+1, & \hspace{7mm} \sigma_i,\sigma_j=A,\\
+1/2, & \hspace{7mm} \sigma_i,\sigma_j=B\quad \mbox{or}\quad \sigma_i\neq \sigma_j.
\end{array} \right.
\end{equation}
For technical reasons, we have introduced in both models a cut-off $r_{ij}=0.5$ for the Lennard-Jones potentials,
below which the potential is repulsive hard-core (i.e., the potential is infinite).

\bigskip

\begin{figure}
\centerline{ \psfig{figure=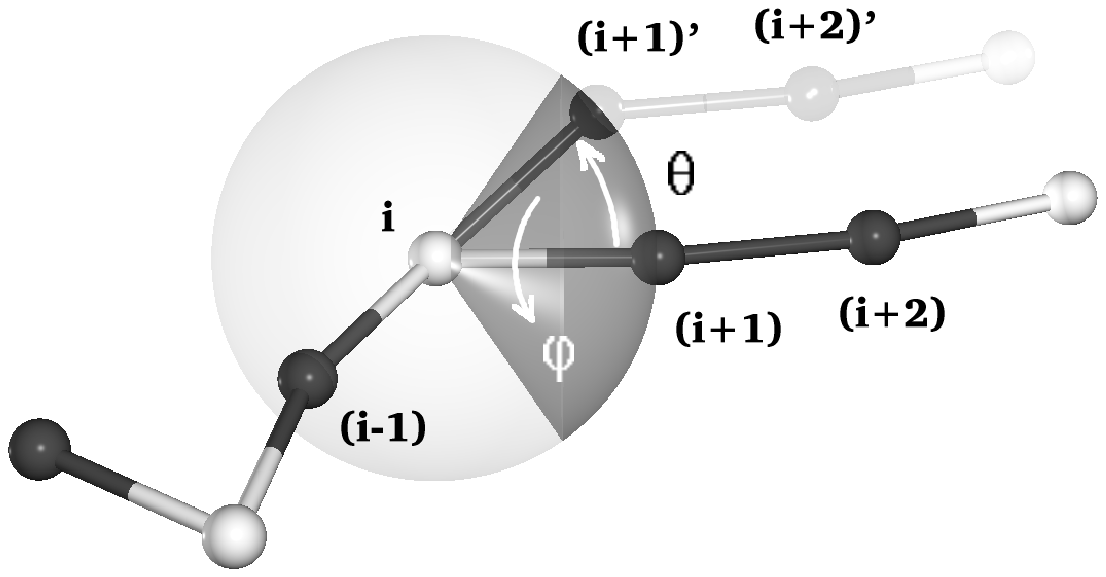}}
\caption{\label{figupdate} Spherical update of the bond vector
between the $i$th and $(i+1)$th monomer.  }
\end{figure}

For updating a conformation, as shown  in Fig.~\ref{figupdate},
 the length of the bonds are fixed ($|{\bf b}_k|=1$, $k=1,\ldots,N-1$). The $(i+1)$th monomer
lies on the surface of a unit sphere centered on   the $i$th monomer.
Therefore, spherical coordinates are the natural choice for calculating the new
position of the $(i+1)$th monomer on this sphere. For the reason of efficiency,
all the points  on the sphere are  not selected for updating, but restricted
 the choice to a
spherical cap with maximum opening angle $2\theta_{\max}$ (the dark area in
Fig.~\ref{figupdate}). Thus, to change the position of the $(i+1)$th monomer to
$(i+1)'$,  we  select the angles $\theta$ and $\varphi$ randomly from the
respective intervals $\cos \theta_{\max} \le \cos \theta \le 1$ and $0\le\varphi \le 2\pi$,
which ensure a uniform distribution of the $(i+1)$th monomer's  positions on the associated
spherical cap.
After updating the position of the
$(i+1)$th monomer, the following monomers in the chain are simply translated
according to the corresponding bond vectors which remained  unchanged in this
 update.
 This is similar to single spin updates in local-update Monte Carlo simulations of the classical Heisenberg model with the
difference that, in addition to local energy changes, long-range
interactions of the monomers are
  to be computed anew after the update,due to changing their
relative position.

\bigskip

\section{The Method}

The multicanonical ensemble  is based on a probability
function in which
the different energies are equally probable.
However, implementation of  the multicanonical algorithm (MUCA)
is not straightforward
because the density of states  $n(E)$ is {\it a priori} unknown.
In practice, one only needs  to know the weights  $\omega$,
\begin{equation}
 w(E) \sim 1/n(E) = \exp [(E-F_{T(E)})/k_BT(E)],
\end{equation}

and these weights are calculated in the
first stage of simulation  process  by an iterative  procedure
in which
the temperatures $T(E)$ are built recursively together with
the microcanonical free energies $F_{T(E)}/k_BT(E)$  up to an
additive constant.
The iterative  procedure is
followed by a long production run based on the fixed $w$'s where
equilibrium configurations are sampled. Re-weighting techniques
(see Ferrenberg and Swendsen~\cite{FeSw88} and literature given in
their second reference)
enable one to obtain Boltzmann averages of various
physical variables  over a wide range of temperatures.

\bigskip

As pointed out above, calculation of the {\it a priori} unknown MUCA weights
is not trivial, requiring an experienced  intervention.
For lattice models, this problem was addressed in a
sketchy way by Berg and \c{C}elik~\cite{BeCe92} and later by
Berg~\cite{Be98}.
An alternative way is to establish an automatic process by
incorporating the statistical errors within the recursion
procedure.

\section{Results and Discussions}

We first carried out canonical (i.e.,
constant $T$) MC simulations of  six different 20-monomer
sequences~\cite{Bachmann}
 at temperatures $T = 0.3,~~0.6~~$ and $2.4$, as well as MUCA test runs
to determine the required energy ranges for each sequence.
The multicanonical weights  were built after $ m = 500 $ recursions during a
long {\it single} simulation, where the multicanonical parameters
were iterated every 10000~sweeps.  After having fixed the MUCA weight
factors, a  production run was carried out with
 $ 5\times 10^{7} $ sweeps.
\bigskip

\begin{figure}%[tb]
\psfig{figure=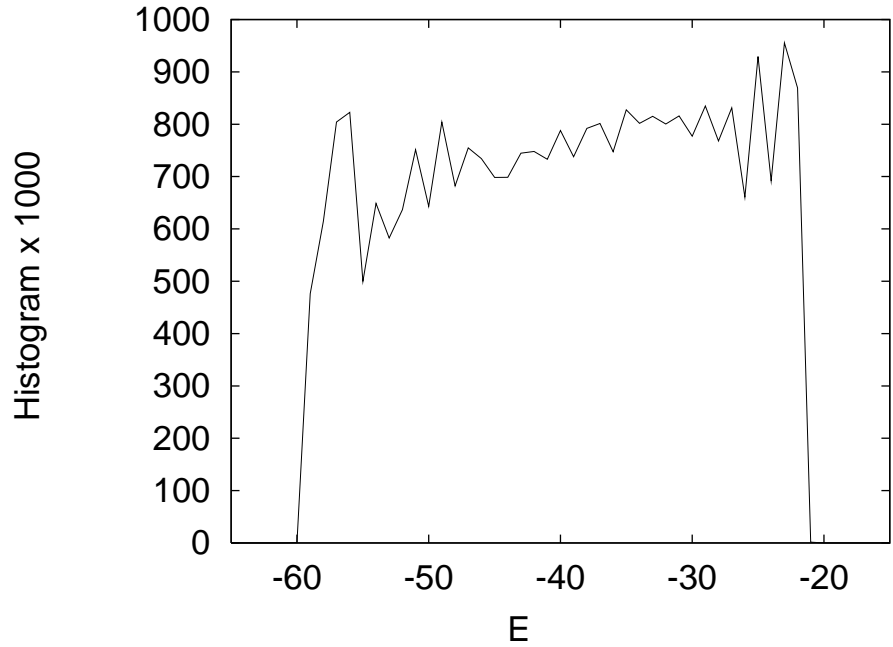}
%\end{figure}
\caption{\label{hist}  Histogram of multicanonical simulations at $T=2.4 K $ for the sequence AAAABBAAAABAABAAABBA.  }
\end{figure}

\noindent Fig.~\ref{hist}  shows, as a sample,  the histograms of the
multicanonical run for the sequence AAAABBAAAABAABAAABBA.
 We can see from the figure
that the whole temperature range, including the hard-to-reach
 low-energy region,
 is equally well sampled.
 Our simulations  with   these
multicanonical parameters
 reveiled the  lowest energy conformation for this sequence (our suspected GEM)
at $ E = -59.105 $.

\bigskip

\begin{figure}%[tb]
\hspace{0.1cm} \psfig{figure=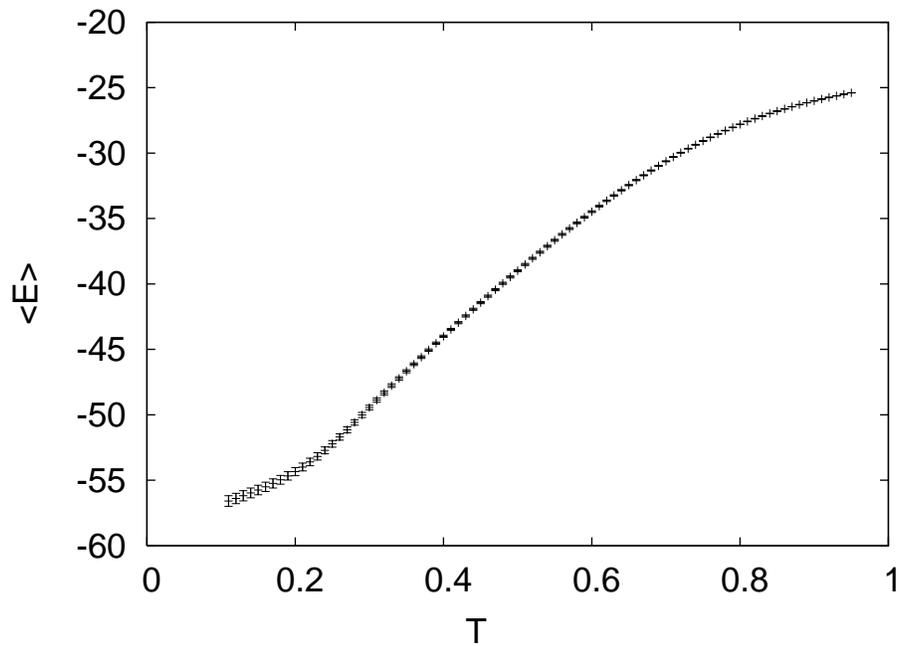}
%\vspace{1pc}
\caption{\label{energy}  The Boltzmann average energy as a function of temperature obtained from multicanonical simulation for the sequence
AAAABBAAAABAABAAABBA. }
\end{figure}

\begin{figure}%[tb]
\hspace{0.1cm} \psfig{figure=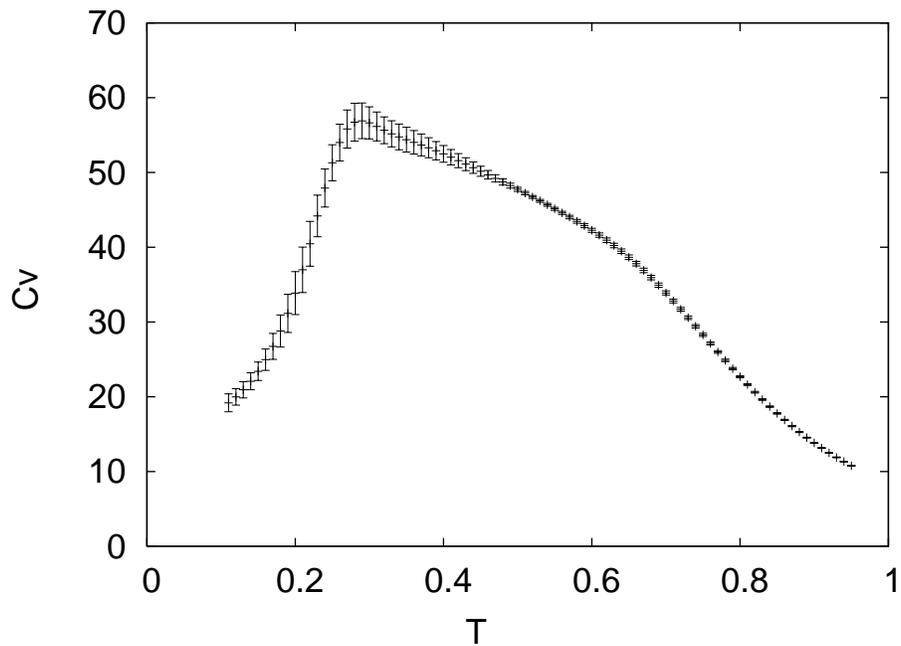}
%\vspace{1pc}
\caption{\label{Cv}  Specific heat obtained from multicanonical simulation as a
 function of temperature. }
\end{figure}

\noindent Fig.~\ref{energy} and Fig.~\ref{Cv} display the energy and the specific
heat vs. temperature for the same sequence. The specific heat possess a peak about the temperature $ T = 0.28  $  , which corresponds to the transition point
where the system undergoes a structural change from random coil to more compact globular state.

\bigskip

\noindent We define an order parameter (OP)~\cite{Hansmann}

\begin{equation}
OP = 1 -\frac{1}{90~n_F} \sum_{i=1}^{n_F} |\alpha_i^{(t)}- \alpha_i^{(RS)}|~,
\end{equation}

\noindent where $n_F$  is the number of bond  and torsion angles,  $\alpha_i^{(RS)}$
and $\alpha_i^{(t)}$ are the bond and torsion  angles of the considered configuration and of the choosen reference state, which is usually taken as the global energy minimum (GEM) state. The difference
 $\alpha_i^{(t)}- \alpha_i^{(RS)}$ is always in the interval
$[-180^{\circ},180^{\circ}]$ , which in turn gives

\begin{equation}
-1 \le ~<O>_T~ \le 1
\end{equation}

\noindent The order parameter may be considered as a measure of coinsedence of the considered conformation with the reference state.

\noindent The  free energy of the system was calculated
by the formula~\cite{Onuchic2}

\begin{equation}
 F (T,OP) = E - TS(OP),
\end{equation}

\noindent and the entropy in the above formula is,
\begin{equation}
 S(OP) = \log H(OP)
\end{equation}

\noindent where $H (OP)$  is the histogram that the system has an order parameter value OP at a fixed temperature $ T$. The
multicanonical algorithm provides a goal sampling of the entire energy
(temperature) range and enables one to determine the histograms $ H$ without problem.

\bigskip

\begin{figure}[!h]
\centerline{\hbox{
\psfig{figure=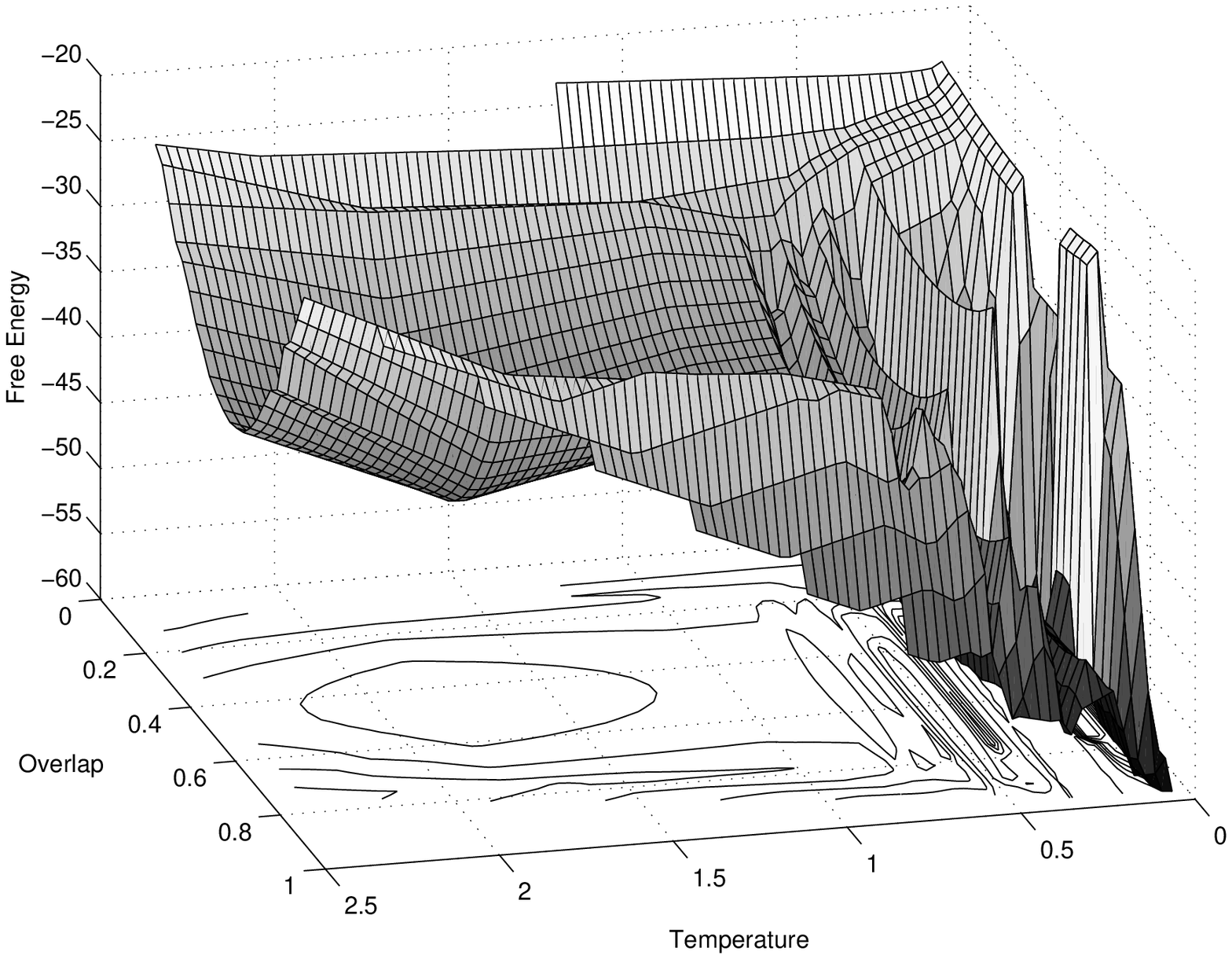,height=14cm,width=16cm} }}
\vspace*{-0.3cm} \caption{\label{free1}  Free energy surface in
the configuration space of sequence  AAAABBAAAABAABAAABBA. }
\end{figure}

\begin{figure}%[tb]
\hspace{0.1cm} \psfig{figure=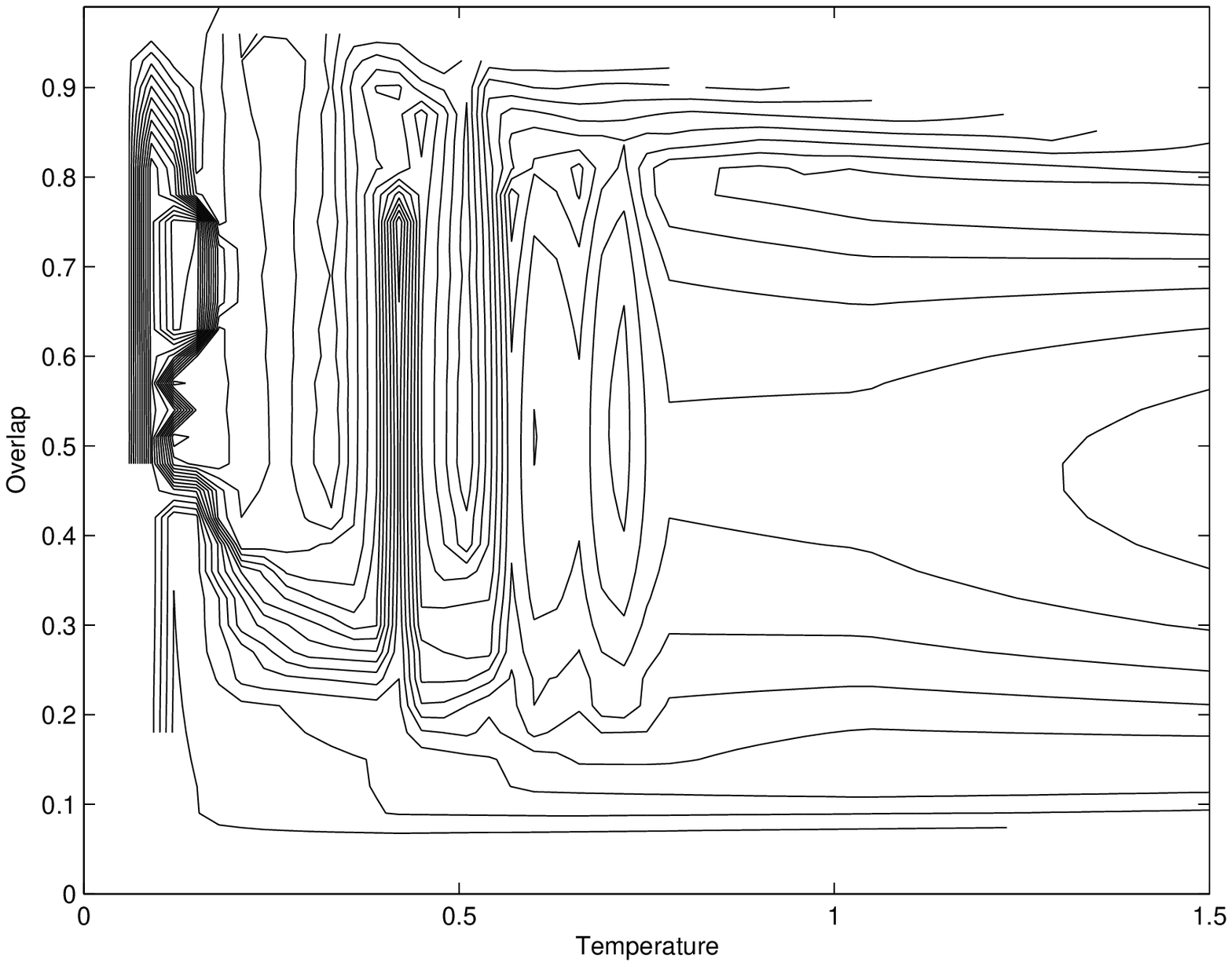}
%\vspace{1pc}
\caption{\label{contour}  Free energy surface contour  in the configuration space of sequence  AAAABBAAAABAABAAABBA . }
\end{figure}

\noindent In Fig.~\ref{free1},  we show
 the free energy surface with respect to the other
parameter $ OP$  and temperature $T$, evaluated by utilizig the
multicanonical technique. The figure displays the free energy
distribution of the created
fifty million conformations of the choosen model protein with respect to the
order parameter and temperature. Instead of potential energy, we have
calculated and displayed the free energies in order to include the entropic
effects, which are more meaningfull in considerations of the folding pathways.
the free energy surface displays a valley structure and clearly
pictures the existing funnel towards the state of global energy minimum.
This topographic picture also allows one to realize at what temperature
the system go through a structural change.
Fig.~\ref{free1}, together with the contour map given in
Fig.~\ref{contour}, enable one to visualize where the system under
consideration  leaves the more structural energy landscape and reaches to a
 funnel towards the lower energy states. This visualization is very helpfull
from point of designing more effective generic simulation algoritms specific
to the system under consideration.

\section{Conclusions}

\noindent We have simulated different 20-monomer sequences of the off-lattice
AB model proteins by utilizig multicanonical ensemble approach and
investigated the free energy surface. The obtained picture serves as a
usefull tool for visualization of the behaviour of considered system in the
configuration space and provide helpfull information in designing more
effective simulation algorithms.

\smallskip
\section{Acknowledgements}
H.A.\ acknowledges support by The Scientific and  Tecnological
Research Council of Turkey (T\"{U}B\.{I}TAK) under the project
number 104T150 and from  The Turkish Academy of Sciences
(T\"{U}BA)
 under the Programme to Reward
Successful Young Scientists.
T.\c{C}. acknowledges the Turkish Academy of Sciences (T\"{U}BA).
Fruitful discussion with G. G\"{o}ko\u{g}lu is acknowledged.

\pagebreak
\newpage

\label{secupdate}

\pagebreak


\pagebreak

\pagebreak

\end{document}